# Neutrinoless double-$\beta$ decay of $^{124}$Sn, $^{130}$Te, and $^{136}$Xe in the Hamiltonian-based generator-coordinate method

C. F. Jiao,[1,2] M. Horoi,[1] and A. Neacsu[1]
[1]*Department of Physics, Central Michigan University, Mount Pleasant, Michigan 48859, USA*
[2]*Department of Physics, San Diego State University, San Diego, California 92182-1233, USA*

We present a generator-coordinate method for realistic shell-model Hamiltonians that closely approximates the full shell model calculations of the matrix elements for the neutrinoless double-$\beta$ decay of $^{124}$Sn, $^{130}$Te, and $^{136}$Xe. We treat not only quadrupole deformations but also the proton-neutron pairing amplitudes as generator coordinates. We validate this method by calculating and comparing spectroscopic quantities with the exact shell model results and experimental data. Our Hamiltonian-based generator-coordinate method produces $0\nu\beta\beta$ matrix elements much closer to the shell model ones, compared to the existing energy-density-functional-based generator-coordinate approaches. The remaining overestimation of $0\nu\beta\beta$ nuclear matrix element suggests that additional correlations may be needed to be taken into account for $^{124}$Sn, $^{130}$Te, and $^{136}$Xe when calculating with the Hamiltonian-based generator-coordinate method. The validation of this method may open the possibility of calculating $0\nu\beta\beta$ matrix element of $^{150}$Nd in a large shell-model space.

## I. INTRODUCTION

The possible detection of $0\nu\beta\beta$ decay is of great importance for unveiling the Majorana character of neutrinos, which would imply an extension to the standard model of particle physics. Furthermore, if this transition is observed, the measurement of its decay rates can provide information about the absolute neutrino mass scale and mass hierarchy, if one can obtain accurate values of the underlying nuclear matrix elements (NMEs) governing the $0\nu\beta\beta$ decay [1]. At present, NMEs given by various nuclear models show significant variation up to a factor of 3 [2,3]. These large uncertainties in the calculated NME preclude not only a definitive choice and amount of material that is required in expensive experiments, but also the inference of accurate values of absolute neutrino masses once a $0\nu\beta\beta$ decay rate is known [2]. Reducing the uncertainty in the matrix elements would be crucial for planning the expensive and complicated $\beta\beta$ experiments that are anticipated.

Several nuclear structure methods have been employed to calculate the $0\nu\beta\beta$ NMEs, including shell model (SM) [4–12], interacting boson model (IBM) [13–15], quasiparticle random phase approximation (QRPA) [16–27], and generator coordinate method (GCM) [28–32]. Both QRPA and GCM can incorporate energy density functional (EDF) theory [23,28–31]. The most obvious feature of EDF-based GCM calculations is that the $0\nu\beta\beta$ matrix elements are much larger than those of the SM. The overestimation could be the result of missing important high-seniority correlations in both the GCM ansatz and in the EDF itself [33]. Adding correlations to EDF without having genuine operators to relate to, however, may lead to inconsistencies and unexpected results [34]. In contrast, SM exactly diagonalizes an effective nuclear Hamiltonian ($H_{\text{eff}}$) in a valence space, which means that SM considers all possible correlations around the Fermi surface that can be induced by $H_{\text{eff}}$. However, this careful treatment of correlations restricts SM calculations to relatively small configuration spaces. Most SM calculations of $0\nu\beta\beta$ matrix elements are performed in a single harmonic oscillator shell, and hence may be unable to fully capture collective correlations [3].

One way to overcome some drawbacks of EDF-based GCM is to abandon the focus on functionals and return to Hamiltonians. Starting from an effective Hamiltonian allows us to include the important shell model correlations in a self-consistent way. In addition, an effective Hamiltonian is usually originated from a bare nucleon-nucleon interaction, and then adapted to a certain configuration space through many-body perturbation theory [35]. This means that the Hamiltonian-based GCM can be extended to larger valence space. The Hamiltonian-based GCM may thus be able to include both the important correlations of SM and large configuration spaces of EDF-based methods, without the drawbacks of either approach.

An important question to answer is which correlations are most relevant to $0\nu\beta\beta$ matrix elements, but are missing in the EDF-based GCM calculations. Both QRPA and SM calculations suggest that isoscalar pairing correlations quench the Gamow-Teller part of $0\nu\beta\beta$ matrix elements significantly [36–38], while the implementation of isoscalar pairing for EDF has not yet been developed. Recently, we developed a GCM code that works with realistic effective Hamiltonians, and we tested it on the calculations of $0\nu\beta\beta$ NMEs for $^{48}$Ca and $^{76}$Ge within both a single shell and two major shells [39]. To fully address the effect of isoscalar pairing and quadrupole correlations on $0\nu\beta\beta$ decay, we treat the isoscalar pairing amplitude, in addition to quadrupole shape fluctuations, as the generator coordinates in the GCM approach.

Recent interest in $^{124}$Sn, $^{130}$Te, and $^{136}$Xe for $0\nu\beta\beta$-decay experiments [40,41] presents a pressing need for accurate calculations of the $0\nu\beta\beta$ NMEs associated with these nuclei. In addition, investigations of the $0\nu\beta\beta$ decay for these nuclei could be a crucial step towards the study of heavier $0\nu\beta\beta$ candidates (e.g., $^{150}$Nd) in extremely large model spaces that are beyond the current capability of SM codes. Therefore, we extend our Hamiltonian-based GCM to the calculation of spectroscopic quantities as well as the $0\nu\beta\beta$ NMEs for $^{124}$Sn, $^{130}$Te, and $^{136}$Xe. This paper is organized as follows: Section II presents a brief overview of the $0\nu\beta\beta$ matrix elements and of the GCM that works with multi-shell effective Hamiltonians. Section III presents the detailed study with the Hamiltonian-based GCM for $^{124}$Sn, $^{130}$Te, and $^{136}$Xe, including the calculated ground-state energies, level spectra, $E2$ transition strength, occupancies of valence orbitals, and the analysis of $0\nu\beta\beta$ NMEs. Finally, a summary with conclusions is presented in Sec. IV.

## II. THE MODEL

In the closure approximation, we can write the $0\nu\beta\beta$ decay matrix element in terms of the initial and final ground states and a two-body transition operator. If the decay is produced by the exchange of a light-Majorana neutrino with the usual left-handed currents, the NME can be given by

$$M^{0\nu} = M^{0\nu}_{GT} - \frac{g_V^2}{g_A^2} M^{0\nu}_F + M^{0\nu}_T \quad (1)$$

where GT, F, and T refer to the Gamow-Teller, Fermi and tensor parts of the matrix elements. The vector and axial coupling constants are taken to be $g_V = 1$ and $g_A = 1.254$, respectively. We modify our wave functions at short distances using a Jastrow short-range correlation (SRC) function with CD-Bonn parametrization [21]. A detailed definition of the form of the matrix element can be found in Ref. [20].

To compute the $0\nu\beta\beta$ matrix element in Eq. (1), one needs initial and final ground states $|I\rangle$ and $|F\rangle$, which can be provided by GCM. We use a shell model effective Hamiltonian ($H_{\text{eff}}$) in our approach. In an isospin scheme, $H_{\text{eff}}$ can be written as a sum of one- and two-body operators:

$$H_{\text{eff}} = \sum_a \epsilon_a \hat{n}_a + \sum_{a \leqslant b, c \leqslant d} \sum_{JT} V_{JT}(ab;cd) \hat{T}_{JT}(ab;cd), \quad (2)$$

where $\epsilon_a$ stands for single-particle energies, $V_{JT}(ab;cd)$ stands for two-body matrix elements (TBMEs), $\hat{n}_a$ is the number operator for the spherical orbit $a$ with quantum numbers $(n_a, l_a, j_a)$, and

$$\hat{T}_{JT}(ab;cd) = \sum_{MT_z} A^\dagger_{JMTT_z}(ab) A_{JMTT_z}(cd) \quad (3)$$

is the scalar two-body density operator for nucleon pairs in orbitals $a, b$ and $c, d$ coupled to the quantum numbers $J, M, T$, and $T_z$.

The first step in the GCM procedure is to generate a set of reference states $|\Phi(q)\rangle$ that are quasiparticle vacua constrained to given expectation values $q_i = \langle \mathcal{O}_i \rangle$ for a set of collective operators $\mathcal{O}_i$. Here we take the operators $\mathcal{O}_i$ to be

$$\begin{aligned}
&\mathcal{O}_1 = Q_{20}, \quad \mathcal{O}_2 = Q_{22}, \\
&\mathcal{O}_3 = \tfrac{1}{2}(P_0 + P_0^\dagger), \quad \mathcal{O}_4 = \tfrac{1}{2}(S_0 + S_0^\dagger),
\end{aligned} \quad (4)$$

where

$$\begin{aligned}
Q_{2M} &= \sum_a r_a^2 Y_a^{2M}, \\
P_0^\dagger &= \frac{1}{\sqrt{2}} \sum_l \hat{l} [c_l^\dagger c_l^\dagger]_{000}^{L=0,J=1,T=0}, \\
S_0^\dagger &= \frac{1}{\sqrt{2}} \sum_l \hat{l} [c_l^\dagger c_l^\dagger]_{000}^{L=0,J=0,T=1},
\end{aligned} \quad (5)$$

with $M$ labeling the angular-momentum $z$ projection, $a$ labeling nucleons, and the brackets signifying the coupling of orbital angular momentum, spin, and isospin to various values, each of which has $z$ projection zero. In Eq. (5), the operator $c_l^\dagger$ creates a particle in the single-particle level with an orbital angular momentum $l$, and $\hat{l} \equiv \sqrt{2l+1}$. The operator $P_0^\dagger$ creates a correlated isoscalar pair and the operator $S_0^\dagger$ a correlated isovector $pn$ pair. To include the effect of isoscalar pairing, we start from a Bogoliubov transformation that mixes neutrons and protons in the quasiparticle operators, i.e. (schematically),

$$\alpha^\dagger \sim u_p c_p^\dagger + v_p c_p + u_n c_n^\dagger + v_n c_n. \quad (6)$$

The actual equations in practical calculations sum over single-particle states in the valence space, so that each of the coefficients $u$ and $v$ in Eq. (6) would be replaced by matrices $U$ and $V$, as described in Ref. [42].

We further solve the constrained Hartree-Fock-Bogoliubov (HFB) equations for the Hamiltonian with linear constraints

$$\langle H' \rangle = \langle H_{\text{eff}} \rangle - \lambda_Z(\langle N_Z \rangle - Z) - \lambda_N(\langle N_N \rangle - N) \\
- \sum_i \lambda_i (\langle \mathcal{O}_i \rangle - q_i), \quad (7)$$

where the $N_Z$ and $N_N$ are the proton and neutron number operators, $\lambda_Z$ and $\lambda_N$ are corresponding Lagrange multipliers, the sum over $i$ includes up to three of the four $\mathcal{O}_i$ in Eq. (4), and the other $\lambda_i$ are Lagrange multipliers that constrain the expectation values of those operators to $q_i$. We solve these equations many times, constraining each time to a different point on a mesh in the space of $q_i$.

Once we obtain a set of HFB vacua with various amounts of axial deformation, triaxial deformation, and isoscalar pairing amplitude, the GCM state can be constructed by superposing the projected HFB vacua as:

$$|\Psi^J_{NZ\sigma}\rangle = \sum_{K,q} f^{JK}_{q\sigma} |JMK; NZ; q\rangle, \quad (8)$$

where $|JMK; NZ; q\rangle \equiv \hat{P}^J_{MK} \hat{P}^N \hat{P}^Z |\Phi(q)\rangle$. The $\hat{P}$'s are projection operators that project HFB states onto well-defined angular momentum $J$ and its $z$ component $M$, neutron number $N$, and proton number $Z$ [44]. The weight functions $f^{JK}_{q\sigma}$, where $\sigma$ is simply a enumeration index, can be obtained by solving the Hill-Wheeler equations [44]

$$\sum_{K',q'} \{\mathcal{H}^J_{KK'}(q;q') - E^J_\sigma \mathcal{N}^J_{KK'}(q;q')\} f^{JK'}_{q'\sigma} = 0, \quad (9)$$

where the Hamiltonian kernel $\mathcal{H}^J_{KK'}(q;q')$ and the norm kernel $\mathcal{N}^J_{KK'}(q;q')$ are given by

$$\mathcal{H}^J_{KK'}(q;q') = \langle \Phi(q)|H_{\text{eff}} \hat{P}^J_{KK'} \hat{P}^N \hat{P}^Z |\Phi(q')\rangle,$$
$$\mathcal{N}^J_{KK'}(q;q') = \langle \Phi(q)|\hat{P}^J_{KK'} \hat{P}^N \hat{P}^Z |\Phi(q')\rangle. \quad (10)$$

To solve Eq. (9), we diagonalize the norm kernel $\mathcal{N}$ and use the nonzero eigenvalues and corresponding eigenvectors to construct a set of "natural states." Then, the Hamiltonian is diagonalized in the space of these natural states to obtain the GCM states $|\Psi^J_{NZ\sigma}\rangle$ (see details in Refs. [45,46]). With the lowest $J = 0$ GCM states as ground states of the initial and final nuclei, we can finally calculate the $0\nu\beta\beta$ decay matrix element $M^{0\nu}$ in Eq. (1).

## III. CALCULATIONS AND DISCUSSIONS

We start by using our GCM which employs an effective Hamiltonian in a model space that a SM calculation can be applied to. If the Hamiltonian-based GCM and the SM itself use the same Hamiltonian and the same valence space, the SM can thus be considered as the "exact" solution because it diagonalizes the Hamiltonian exactly. Therefore, comparing the results given by Hamiltonian-based GCM and SM indicates to what extent the GCM can capture the correlations relevant to $0\nu\beta\beta$ matrix elements.

For the calculations of the $0\nu\beta\beta$ decay NMEs of $^{124}$Sn, $^{130}$Te, and $^{136}$Xe, we need a reliable effective Hamiltonian. Currently, the best option is using a fine-tuned effective Hamiltonian for the $jj55$-shell configuration space that compromises the $0g_{7/2}$, $1d_{5/2}$, $1d_{3/2}$, $2s_{1/2}$, and $0h_{11/2}$ orbitals. As a reference Hamiltonian, we use a recently proposed shell-model effective Hamiltonian called the singular value decomposition (SVD) Hamiltonian [47], which has been tested for the range of nuclei of interest to this study [9,10,47]. This Hamiltonian accounts successfully for the spectroscopy, electromagnetic and Gamow-Teller transitions, and deformation of the initial and final nuclei that our calculations involve [9,10]. Although the energy spectra of $^{136}$Xe and $^{136}$Ba produced by the SVD Hamiltonian are not as good as those one gets when using the GCN50:82 Hamiltonian [48], they are still reasonable when compared with experimental data (see Fig. 1 of Ref. [9]). In addition, Ref. [9] used the SVD Hamiltonian to check all other relevant observables, such as the reduced $E2$ transition probabilities, $B(E2: 0^+_1 \to 2^+_1)$, the occupation probabilities, and the Gamow-Teller strengths, which were found again to be in reasonably good agreement with the available experimental data when the standard effective charges and quenching factors are used. Therefore, we assume that our analysis based on the SVD effective Hamiltonian is not unreasonable, and can be used to asses the viability of our GCM approach.

Because these nuclei show no evidence of triaxial deformation, neither theoretically nor experimentally, our GCM calculations use axial quadrupole moment $q_1 \equiv \langle Q_{20}\rangle$, as well as the proton-neutron pairing parameters $q_3 \equiv 1/2\langle P_0 + P_0^\dagger\rangle$ and $q_4 \equiv 1/2\langle S_0 + S_0^\dagger\rangle$ as generator coordinates.

TABLE I. The g.s. energies (in MeV) obtained with the SVD Hamiltonian by using GCM and SM for $^{124}$Sn, $^{124}$Te, $^{130}$Te, $^{130}$Xe, $^{136}$Xe, and $^{136}$Ba.

| Nuclei | GCM | SM |
| --- | --- | --- |
| $^{124}$Sn | −15.733 | −16.052 |
| $^{124}$Te | −23.082 | −24.446 |
| $^{130}$Te | −25.705 | −26.039 |
| $^{130}$Xe | −32.583 | −33.313 |
| $^{136}$Xe | −34.931 | −34.971 |
| $^{136}$Ba | −40.341 | −40.745 |

### A. Ground-state energies and spectra

The first study we perform with the Hamiltonian-based GCM is the comparison of the calculated ground-state (g.s.) energies and the low-lying spectra of $^{124}$Sn, $^{124}$Te, $^{130}$Te, $^{130}$Xe, $^{136}$Xe, and $^{136}$Ba with the SM calculations [9,10]. Table I presents the g.s. energies obtained using the SVD Hamiltonian for both nuclear structure approaches employed. These results are very close to the exact solutions given by SM, implying that our Hamiltonian-based GCM has captured a major part of important correlations in the ground states of these nuclei.

Figure 1 shows the low-lying level spectra of these nuclei, compared to the SM calculations [9,10] and experimental data [43]. Generally, the Hamiltonian-based GCM calculations are in reasonable agreement with the SM results and experimental spectra. As a comparison, the EDF-based GCM gives much higher $2^+$ states for $^{124}$Sn, $^{124}$Te, $^{130}$Te, $^{130}$Xe, $^{136}$Xe, and $^{136}$Ba [30]. We notice that the GCM calculations tend to overestimate the excitation energies of $4^+$ states given by SM. The overestimation could be due to the fact that our GCM calculations exclude the multi-quasiparticle (or called broken-pair) excitation, and this breaks the time-reversal symmetry. The multi-quasiparticle excitations would lower the excited states significantly, especially in the nearly spherical and weakly deformed nuclei. The inclusion of noncollective configurations like multi-quasiparticle configurations would be important for a better description of the low-lying spectra of spherical and weakly deformed nuclei. This is, however, beyond scope of this work.

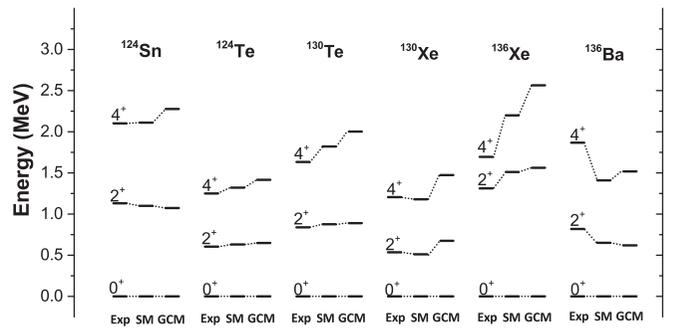

FIG. 1. The calculated low-lying energy levels for $^{124}$Sn, $^{124}$Te, $^{130}$Te, $^{130}$Xe, $^{136}$Xe, and $^{136}$Ba, compared to the exact solutions of SM [9,10] and experimental data [43].

TABLE II. The $B(E2: 0_1^+ \to 2_1^+)$ (in $e^2\text{b}^2$) obtained with the SVD Hamiltonian by using GCM and SM for $^{124}$Sn, $^{124}$Te, $^{130}$Te, $^{130}$Xe, $^{136}$Xe, and $^{136}$Ba, compared to the adopted values [49].

|        | $^{124}$Sn | $^{124}$Te | $^{130}$Te | $^{130}$Xe | $^{136}$Xe | $^{136}$Ba |
|--------|-----------|-----------|-----------|-----------|-----------|-----------|
| GCM    | 0.168     | 0.648     | 0.165     | 0.492     | 0.220     | 0.475     |
| SM     | 0.146     | 0.579     | 0.153     | 0.502     | 0.215     | 0.479     |
| Adopted| 0.162     | 0.560     | 0.297     | 0.634     | 0.217     | 0.413     |

### B. $B(E2) \uparrow$ transitions

For the calculation of the reduced $E2$ transition probability $B(E2: 0_1^+ \to 2_1^+)$, we use the canonical effective charges $e_n^{\text{eff}} = 0.5e$, $e_p^{\text{eff}} = 1.5e$ for $^{130}$Te/Xe and $^{136}$Xe/Ba. For $^{124}$Sn/Te, we use $e_n^{\text{eff}} = 0.88e$ and $e_p^{\text{eff}} = 1.88e$, which are suggested for tin isotopes with no protons in the valence space in Refs. [10,47]. The results are presented in Table II, in comparison with the SM calculated values [9,10] and the experimentally adopted ones [49]. We can see a very good agreement between the Hamiltonian-based GCM and the SM calculations. Both GCM and SM calculations reproduce well the adopted values with only a slight underestimation for $^{130}$Te and $^{130}$Xe. The underestimation may indicate the need of adjusting the effective charges.

### C. Occupation probabilities

It is known that the $0\nu\beta\beta$ NMEs are sensitive to the occupancies of valence neutron and proton orbitals [22]. To further verify how suitable the Hamiltonian-based GCM is in describing the nuclear structure and $0\nu\beta\beta$ decay aspects of the nuclei involved, we also calculate the neutron vacancies for $^{124}$Sn, $^{124}$Te, $^{136}$Xe, and $^{136}$Ba, as well as the proton occupancies for $^{124}$Te and $^{136}$Ba. Our results are presented in Fig. 2, and are compared to the SM calculations reported in Figs. 2–4 of Ref. [9] and Fig. 2 of Ref. [10]. The occupation probabilities of the $1d_{5/2}$ and $1d_{3/2}$ orbitals are summed up and presented as $1d$, which is similar to the procedure used in the references that provides the experimental data for comparison. The occupancies obtained with our GCM calculations are close to the values calculated by SM. The calculated occupancies, combined with the g.s. energies, level spectra, and $B(E2: 0_1^+ \to 2_1^+)$ mentioned above, leads us to consider that Hamiltonian-based GCM is suitable for the description of the relevant nuclear structure aspects for these initial and final nuclei involved in the $0\nu\beta\beta$ decay.

### D. Analysis of the $0\nu\beta\beta$ nuclear matrix elements for $^{124}$Sn, $^{130}$Te, and $^{136}$Xe

The validation of the Gamow-Teller strength distributions and $2\nu\beta\beta$ decay NME is relevant for a good description of $0\nu\beta\beta$ decay rates. Both of them require a theoretical calculation of an odd-odd nucleus (e.g., $^{136}$Cs for $^{136}$Xe), which is the daughter nucleus of the Gamow-Teller transition of the parent (e.g., $^{136}$Xe) and the intermediate nucleus in the $2\nu\beta\beta$ decay. However, the calculations of odd-odd nuclei in the framework of GCM associated with angular-momentum and particle-number projection have not been fully developed. This is because (i) even at the mean-field level, such as the HFB approach, odd and odd-odd nuclei are numerically difficult to calculate, and therefore for the ground states one must try several spins or parity; (ii) the blocked structure of the wave function results in the breaking of the time-reversal symmetry, and hence the triaxial projections have to be performed. Very recently, the GCM has been extended to include studies of odd-even nuclei, which have been studied using the Skyrme [50] and Gogny forces [51], but a realistic GCM calculation of odd-odd nuclei seems to be still out of reach. Therefore, the Gamow-Teller strength distributions and $2\nu\beta\beta$ decay NMEs are not discussed in this paper.

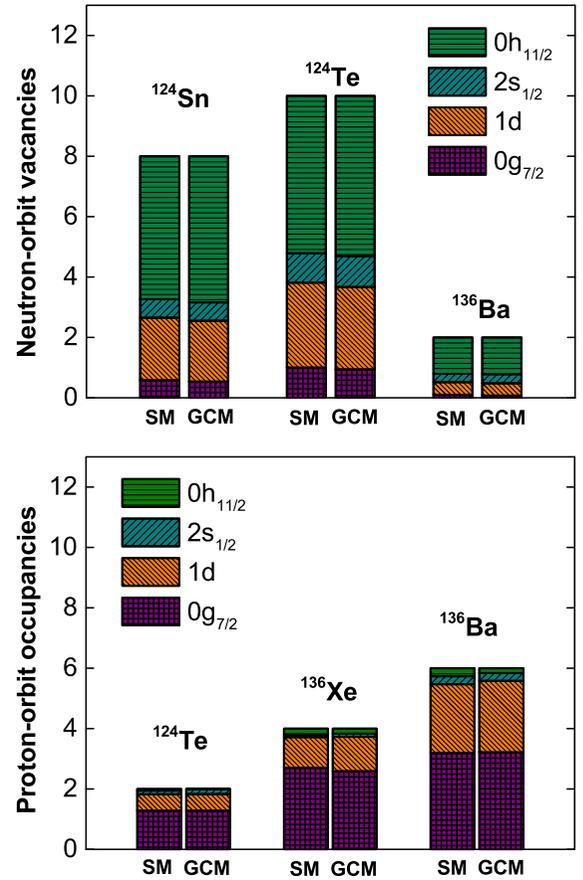

FIG. 2. The calculated vacancies of valence neutron orbitals for $^{124}$Sn, $^{124}$Te, and $^{136}$Ba, as well as the calculated occupancies of valence proton orbitals for $^{124}$Te $^{136}$Xe, and $^{136}$Ba.

We calculate the matrix elements of $^{124}$Sn, $^{130}$Te, and $^{136}$Xe using our GCM approach. Recent interest in these nuclei for $0\nu\beta\beta$-decay experiments [40,41] presents a pressing need for accurate calculations of the $0\nu\beta\beta$ NMEs to guide the experimental effort.

Figure 3 illustrates our calculated $0\nu\beta\beta$ NMEs, compared to the values given by the SM and the EDF-based GCM employing nonrelativistic Gogny D1S force and relativistic

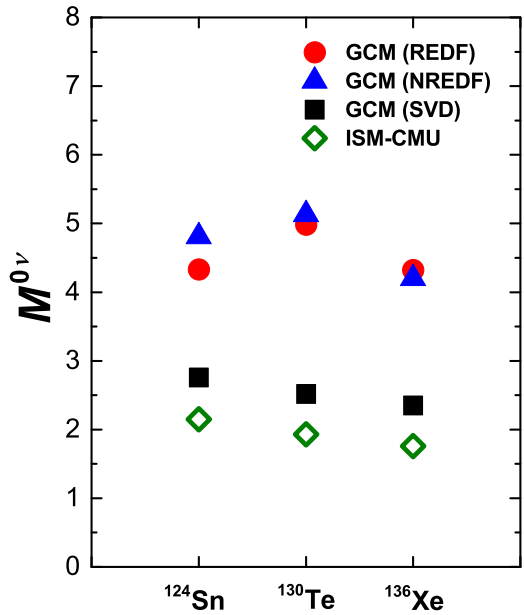

FIG. 3. Calculated $0\nu\beta\beta$ NMEs, compared to the values given by SM (denoted by "ISM-CMU") [9,10], EDF-based GCM employed non-relativistic Gogny D1S force (denoted by "NREDF") [29] and relativistic PC-PK1 force (denoted by "REDF") [30].

PC-PK1 force. It can be seen that previous GCM calculations produce NMEs which are more than two times larger than the SM ones. In contrast, with our Hamiltonian-based GCM calculations we obtain NMEs that are about only 30% larger than the SM results, significantly reducing the long-debated discrepancy between previous GCM and SM predictions.

To further understand this 30% overestimation given by our GCM calculations, the values for the $0\nu\beta\beta$ decay NMEs of $^{124}$Sn, $^{130}$Te, and $^{136}$Xe are listed in Table III, where we show the Gamow-Teller, the Fermi, and the tensor contributions. Generally, the Fermi and tensor parts of NMEs present good agreement between our GCM calculations and SM calculations, while the Gamow-Teller part of NMEs are noticeably larger in our GCM results, resulting in the 30% overestimation in the total $0\nu\beta\beta$ NMEs.

The analysis of the $0\nu\beta\beta$ NME is extended by looking at the decomposition of the NMEs over the angular momentum $I$ of the proton (or neutron) pairs (see Eq. (B4) in Ref. [52]), called $I$-pair decomposition. In this case, the NME can be written as $M_\alpha = \sum_I M_\alpha(I)$, where $M_\alpha(I)$ represent the contributions from each pair-spin $I$ to the $\alpha$ part of the NME. To analyze the deviation between the $M^{0\nu}_{GT}$ given by GCM and SM, Fig. 4 presents the $I$-pair decomposition for the Gamow-Teller part of our calculated $0\nu\beta\beta$ NMEs, compared to the one calculated by SM [9,10]. The bars in Figs. 4 can be added directly to get the Gamow-Teller part of NMEs. As we can see, the dramatic cancellation between the $I = 0$ and $I = 2$ contributions shown by the SM calculations is reproduced perfectly by our GCM approach. However, SM calculations give more negative contributions with $I \geqslant 4$, which further reduce the Gamow-Teller NMEs. In contrast, our GCM approach can barely produce any contributions with $I \geqslant 4$.

TABLE III. The NMEs obtained with SVD Hamiltonian by using GCM and SM for $^{124}$Sn, $^{130}$Te, and $^{136}$Xe. The SM results are taken from Refs. [9,10]. CD-Bonn SRC parametrization was used.

|  |  | $M^{0\nu}_{GT}$ | $M^{0\nu}_F$ | $M^{0\nu}_T$ | $M^{0\nu}$ |
|---|---|---|---|---|---|
| $^{124}$Sn | GCM | 2.48 | −0.51 | −0.03 | 2.76 |
|  | SM | 1.85 | −0.47 | −0.01 | 2.15 |
| $^{130}$Te | GCM | 2.25 | −0.47 | −0.02 | 2.52 |
|  | SM | 1.66 | −0.44 | −0.01 | 1.94 |
| $^{136}$Xe | GCM | 2.17 | −0.32 | −0.02 | 2.35 |
|  | SM | 1.50 | −0.40 | −0.01 | 1.76 |

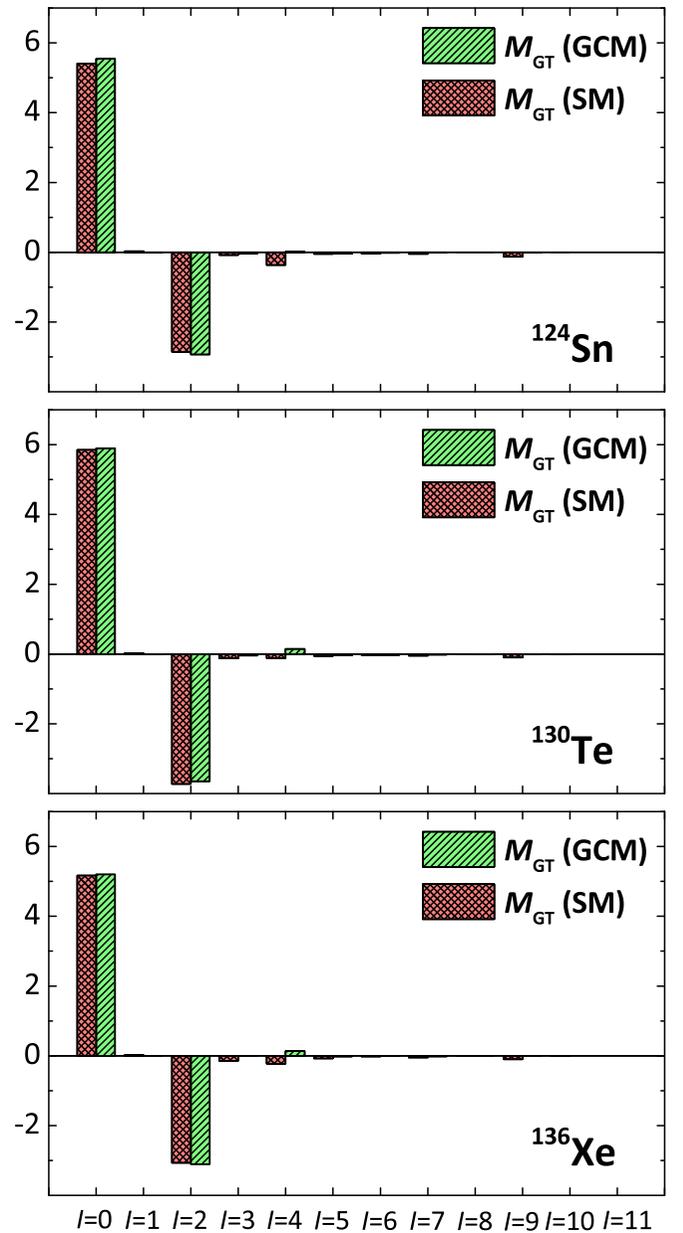

FIG. 4. $I$-pair decomposition: contributions to the Gamow-Teller matrix elements for the $0\nu\beta\beta$ decay of $^{124}$Sn, $^{130}$Te, and $^{136}$Xe from the configurations when two initial neutrons and two final protons have a certain total spin $I$, compared to SM [9,10] calculations. CD-Bonn SRC parametrization was used.

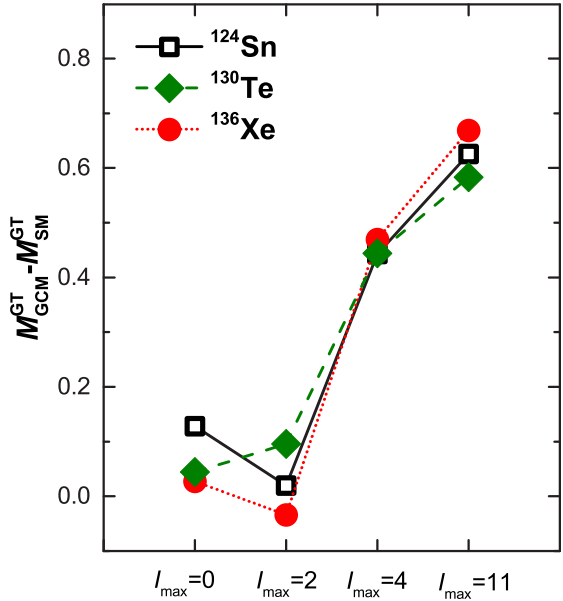

FIG. 5. The differences of Gamow-Teller part of NMEs between our GCM and SM calculations against the pair-spin $I$ for $^{124}$Sn, $^{130}$Te, and $^{136}$Xe.

Figure 5 visualizes the differences in Gamow-Teller NMEs between our GCM and SM calculations against the pair-spin $I$, which can help us to identify where the differences mainly come from. If we only include the $I \leqslant 2$ contributions, the Gamow-Teller NMEs obtained by our GCM approach are very close to the ones given by SM in all three nuclei involved. However, if the $I \leqslant 4$ contribution are taken into account, the differences are noticeably increased to about 0.45 for all the candidate nuclei. The inclusion of all possible pair-spin $I$ contributions would increase these differences even further. Apparently, the overestimation of Gamow-Teller NMEs is associated with those large-$I$-pair contributions, which may correspond to collective or noncollective correlations that are excluded from current GCM calculations.

Therefore, the deviation between our current Hamiltonian-based GCM and SM results may be related to the lack of some correlations, which become important in $^{124}$Sn, $^{124}$Te, $^{130}$Te, $^{130}$Xe, $^{136}$Xe, and $^{136}$Ba. Since these nuclei are all near spherical or weakly deformed, one can expect that the noncollective correlations, for example, quasiparticle excitations, may overcome the collective correlations. Currently, the reference states that the GCM method employs are HFB states imposed by the time-reversal symmetry, which exclude any multi-quasiparticle configurations. It would be of great interest if we could treat the quasiparticle excitation as an additional generator coordinate in the future. It could improve the description of $0\nu\beta\beta$ decay NME for these nuclei.

## IV. SUMMARY

In this paper, we present a GCM calculation based on effective shell model Hamiltonians for the $0\nu\beta\beta$ decay NMEs of $^{124}$Sn, $^{130}$Te, and $^{136}$Xe in the $jj55$ model space that compromises the $0g_{7/2}$, $1d_{5/2}$, $1d_{3/2}$, $2s_{1/2}$, and $0h_{11/2}$ orbitals. We use the SVD effective Hamiltonian that was fine-tuned to describe the experimental data. To ensure the reliability of the results, we perform the Hamiltonian-based GCM calculations of the ground-state energies, low-lying level spectra, and occupancies of valence neutron and proton orbitals. These are compared with the SM results obtained by exactly diagonalizing the same effective Hamiltonian. Our results are in reasonable agreement with the values obtained with the shell model. We also provide a detailed analysis of $0\nu\beta\beta$ decay NMEs for $^{124}$Sn, $^{130}$Te, and $^{136}$Xe. Our Hamiltonian-based GCM produces $0\nu\beta\beta$ decay NMEs that are about 30% larger than the ones obtained by SM, significantly reducing the large deviation between previous GCM and SM predictions. By checking the decomposition of the NMEs over the angular momentum $I$ of the proton or neutron pairs, we find that the remaining 30% overestimation of $0\nu\beta\beta$ decay NMEs may be associated with the exclusion of some noncollective correlations.

Furthermore, the validation of Hamiltonian-based GCM calculation for $^{124}$Sn, $^{130}$Te, and $^{136}$Xe opens the possibility to study the $0\nu\beta\beta$ NME of $^{150}$Nd with realistic effective interactions, by including the most important correlations. For $^{150}$Nd and $^{150}$Sm, a larger valence space (e.g., the shell closures from $N = Z = 50$ to 126) is required. However, the number of states for $A = 150$ nuclei in this space is too large for exact diagonalization in the conventional SM. In contrast, the Hamiltonian-based GCM can be easily extended to this model space. There are at least two ways to obtain a reasonable effective Hamiltonian in this larger model space: (i) build a separable collective Hamiltonian including the monopole term, pairing term, quadrupole-quadrupole term, and spin-isospin term, following the work of Dufour and Zuker [53]; (ii) perhaps more difficult, but more important, is an implementation of valence-space Hamiltonians derived from *ab initio* approaches, such as in-medium similarity renormalization group (IM-SRG) [54] or coupled cluster (CC) [55] methods. Recent work [56] has extended the reach of valence-space IM-SRG in one single shell to essentially all light- and medium-mass nuclei. The extension of the *ab initio* valence-space Hamiltonian in multiple shells by using our GCM approach would be a desirable next step.

## ACKNOWLEDGMENTS

C.F.J. thanks J. Engel, J. M. Yao, and L. J. Wang for helpful discussions and advice. Support from the U.S. Department of Energy Topical Collaboration Grant No. DE-SC0015376 is acknowledged.